\documentclass[12pt]{article}
\newtheorem{Def}{Definition}
\usepackage{graphicx}

\begin{document}

\pagestyle{empty}

\title{Problems of the CASCADE Protocol and Renyi Entropy
Reduction in Classical and Quantum Key Generation}

\author{Koichi Yamazaki$^{1,2}$, Ranjith Nair$^{2}$,
        and Horace P. Yuen$^2$\\
$^{1}$Department of Media-Network Sciences, Tamagawa University,\\
6-1-1, Tamagawa gakuen, Machida, Tokyo, 194-8610, Japan\\
         $^{2}$Department of Electrical Engineering and Computer Science,\\
Northwestern University, Evanston, IL 60208, USA} \maketitle

\begin{abstract}
It is shown that the interactive error correction protocol
`CASCADE' should be analyzed taking the correlation between passes
and finite length of sequence into account. Furthermore we mention
some problems in quantifying the reduction of Renyi entropy by
information announced during the error correction process.
\end{abstract}

\setcounter{section}{0}

\section{Introduction}

The CASCADE protocol for error correction was studied by Brassard
and Salvail \cite{BS94} who gave asymptotic estimates of its
success probability and the number of information bits leaked
during protocol execution. In a practical implementation of
CASCADE, with a fixed number of passes and a finite length
sequence, we show that calculation of the success probability and
number of information bits leaked are very involved problems which
have not been properly addressed so far. Even the estimates in
\cite{BS94} are made under what may be called an `independent
pass' approximation because the correlation between operations
taking place at different passes of the protocol has been
neglected. On the implementation side, we study practically
important parameters such as the communication complexity and the
round-trip delay time of CASCADE.

For cryptographic purposes, in both classical and
quantum protocols, one is interested in the Renyi information
leakage from public discussion during the
error correction process. This is because the standard
privacy amplification theorem \cite{BBCM95} requires the Renyi
entropy before privacy amplification to be known.
Such estimates of Renyi entropy reduction, especially
for the case of identical individual attacks by Eve,
have been studied by Cachin and Maurer \cite{CM97}, who
claimed to show that for linear error-correcting codes,
the Renyi entropy reduction is roughly equal to the
number of announced bits. We explain why this claim
is incorrect. However, we suggest that covering the
announced bits with a pre-shared secret key would result
in no Renyi entropy reduction and can be used as
long as the net key generation rate remains positive.
For CASCADE, for which the Renyi entropy reduction
has not been studied, we suggest a similar procedure.
In doing so, we find that, unlike the case of
linear error-correcting codes, there is still an entropy
reduction that is difficult to quantify.

\section{CASCADE PROTOCOL}
\subsection{`Independent Pass' and `Infinite-length Sequence'  Approximation}
The CASCADE protocol proceeds in several passes in which
block-parity conparison and binary-search-type error correction
are executed. CASCADE uses correlation between the sequences of
passes, so that it performs better as to the number of announced
bits during the error correction process than the BBBSS
protocol\cite{BBBSS92} which is also based on parity check and
binary search, but does not use the correlation of sequences.
Upper bounds on the number of announced bits were derived in
\cite{BS94}. The number of announced bits seems to depend heavily
on error patterns and permutation functions used to produce
sequences of all passes. Furthermore the process is nonlinear.
Therefore, the expected value should be obtained by averaging the
number of announced bits of all specific error patters and
permutation functions. However, the upper bound on the number of
announced bits was derived by calculating expected amount of
announced bits pass by pass. That is, the correlation of sequences
was not considered. We performed computer simulation 1,000,000
times to analyze the performance of CASCADE. The number of
announced bits during the error correction process obtained by the
computer simulation is shown in Tab.\ref{tab.exbit}. The upper
bound derived in \cite{BS94} is also shown in the bracket. It is
found that the upper bound is very close to the simulation
brackets. In the case that the private channel error probability
$\varepsilon$ is 0.05 and 0.1, the simulation results are larger
than the upper bound. This indicates that a nonrigorous method was
used to derive the upper bound.

In \cite{BS94}, the size of a block for each pass was designed so
that the number of errors contained in a block of the initial pass
reduces to less than half as passes proceed. However, the success
probability was discussed in \cite{BS94} on the basis of 100
empirical tests in which all errors were corrected at the end of
the protocol. Analysis of the success probability is very
complicated. Only a slight difference of error positions may
determine whether the protocol succeeds or fails. We have not
derived the probability that CASCASE protocol succeeds. In this
situation, the results of a large number of computer simulations
may help us to understand its properties. The protocol failure
probability obtained by averaging the results of 1,000,000
computer simulation is shown in Tab.\ref{tab.failure}. It is found
that the protocol failure probability reduces as the sequence
length becomes longer. Furthermore it is found that the failure
probability decreases as the error probability of the private
channel increases. According to our computer simulation, we found
that the average number of errors remaining after pass 2 is almost
the same in the range from 0.3 to 0.37. As the error probability
of the private channel decreases, block sizes become longer and
the number of blocks decreases, so that the probability that two
or more of the remaining errors move to the same block in the next
pass becomes large. Then, the protocol failure probability
increases. From this result, the protocol failure probability
should be analysed for sequences with finite length and asymptotic
limits are not reliable.

\begin{table}[h]
\caption{\label{tab.exbit} \textbf{Number of Announced Bits during
Error Correction}}
\begin{center}
\begin{tabular}{|c|c|c|c|c|}

\hline
$n$&$\varepsilon=0.01$&0.05&0.1&0.15\\
\hline
100&13.43(9.33)&35.37(33.14)&59.3(57)&77.56(82.4)\\
\hline
200&20.64(18.66)&69.3(66.29)&117.39(114)&153.86(164.8)\\
\hline
500&45.93(46.64)&170.12(165.71)&288.67(285)&384.76(412)\\
\hline
1000&90.99(93.29)&339.01(331.43)&576.53(570)&768.19(824)\\
\hline
2000&182.16(186.58)&677.25(662.86)&1151.7(1140)&1536.02(1648)\\
\hline
5000&454.38(466.44)&1692.09(1657.14)&2878.41(2850)&3839.52(4120)\\
\hline
10000&906.18(932.88)&3382.11(3314.29)&5753.93(5700)&7678.68(8240)\\
\hline

\end{tabular}
\\
$\varepsilon$: error probability of private channel, $n$: sequence length.\\
The upper bound \cite{BS94} is given in parentheses.

\end{center}
\end{table}

\begin{table}[h]
\caption{\label{tab.failure} \textbf{Protocol Failure Probability
of CASCADE}}
\begin{center}
\begin{tabular}{|c|c|c|c|c|c|c|c|}
\hline
$\varepsilon$ &$n$=100&200&500&1000&2000&5000&10000\\
\hline
0.01&0.023603&0.056623&0.078354&0.044117&0.011357&0.002677&0.001005\\
\hline
0.05&0.072621&0.038288&0.005492&0.001657&0.000355&0.000075&0.000066\\
\hline
0.1&0.032161&0.009446&0.001129&0.000349&0.000119&0.000009&0.000011\\
\hline
0.15&0.016245&0.006394&0.000415&0.000158&0.000039&0&0.00001\\
\hline
\end{tabular}\\
$\varepsilon$: error probability of private channel, $n$: sequence
length.
\end{center}
\end{table}

\begin{table}[h]
\caption{\label{tab.com} \textbf{Number of Round Trips}}
\begin{center}
\begin{tabular}{|c|c|c|c|c||c|c|c|c|}
\hline
&\multicolumn{4}{c||}{CASCADE}&\multicolumn{4}{c|}{BBBSS}\\
\hline
$n$&$\varepsilon$0.01&0.05&0.1&0.14&0.01&0.1&0.1&0.15\\
\hline
500&26.12&58.66&81.95&105.21&26.08&36.04&38.77&38.57 \\
\hline
1000&42.70&107.41&154.89&202.34&33.26&44.16&47.60&47.56\\
\hline
2000&74.86&205.09&300.89&396.46&42.47&54.17&57.87&57.17 \\
\hline
5000&170.21&498.91&739.75&979.01&56.20&68.58&72.29&71.17 \\
\hline
10000&329.23&987.53&1470.49&1949.51&67.42&80.14&82.73&82.27\\
\hline
\end{tabular}
\end{center}
\end{table}

\subsection{Communication Complexity}
For interactive error correction like CASCADE, a round-trip is
necessary for comparing block and sub-block parities. From a
practical viewpoint, time for the round-trips should be considered
in evaluating the net key generation rate. The number of
round-trips is shown in Tab.\ref{tab.com} for CASCADE and BBBSS
protocol which does not use the correlation between
sequences\cite{BBBSS92}. It is found in Tab.\ref{tab.com} that the
number of the round-trip increases as error probability of the
private channel becomes large and the sequence length becomes
longer, and it is almost proportionally to the sequence length in
CASCADE, while that of BBBSS protocol is almost constant. The
difference comes from the fact that each error has to be corrected
one by one after pass two in CASCADE, while errors can be
corrected simultaneously in each pass in BBBSS protocol. It is
found that as sifted key generation rate over a private channel
increases, error correction process would dominate the net key
generation rate, especially for CASCADE.

\section{REDUCTION OF RENYI ENTROPY BY ERROR CORRECTION}
The amount of information leaked during the error correction
process would reduce Eve's Renyi entropy about the legitimate
users' sequence. The precise estimation of the amount of this
reduction is very important for evaluating the key generation
rate. It was shown in Th.~9 of \cite{CM97} that in the case where
the raw key is generated by many independent repetitions of a
random experiment, each bit leaked during the error correction
process reduces Eve's Renyi entropy by only about one. The
properties of the  so-called $\epsilon$-strongly typical sequences
defined below are used to prove this result.

\begin{Def}[$\epsilon$-strongly typical set\cite{CM97}]
Let $X$ be a random variable distributed according to $P_X$ over some finite
set \({\cal X}\) where it is assumed that $P_X(x)>0$ for all $x\in$\({\cal X}\).
Let $x^n=[x_1,...,x_n]$ be a sequence of $n$ digits of \({\cal X}\) and define $N_a(x^n)$
to be the number of occurrences of the symbol $a\in$\({\cal X}\) in the sequence $x^n$.
 A sequence $x^n\in$\({\cal X}\)$^n$ is called $\epsilon$-strongly typical if and only if
$
(1-\epsilon)P_X(a)\leq \frac{N_a(x^n)}{n}\leq(1+\epsilon)P_X(a)
$
for all $a\in$\({\cal X}\).
\end{Def}

It was claimed that the occurrence probability of each sequence in
the $\epsilon$-strongly typical set is asymptotically identical,
and this property was used to prove Th.~9 in \cite{CM97}. This
claim is wrong, however -- the ratio of the maximum occurrence
probability of a sequence in the $\epsilon$-strongly typical set
to the minimum one increases as the sequence length becomes
longer\cite{G01}. For the binary case, assuming that
$P_X(0)=p(\leq 1/2)$, the ratio is
$(\frac{1-p}{p})^{2np\epsilon}$. As an example, let us consider
the case that $n$=100,000, $P_X(0)$=0.1, $\epsilon$=0.01. The
minimum occurrence probability $P_{X^n}^{\rm
min}(x^n)=2.52\times10^{-14214}$, the maximum $P_{X^n}^{\rm
max}(x^n)= 1.78\times10^{-14023}$, then the ratio $P_{X^n}^{\rm
max}(x^n)/P_{X^n}^{\rm min}(x^n)=7.06\times10^{190}$. In this
case, the probability $Pr\left[X^n\in S^n(\epsilon)\right]$ that
the sequence is in a $\epsilon$-strongly typical set
$S^n(\epsilon)$ is 0.711. Therefore, Th.~9 does not apply to this
case. The amount of information additional to that predicted by
Th.~9 can be obtained by Th.~6 of \cite{CM97}. We find the upper
bound on the reduction of Eve's Renyi entropy is 2485 bits more
than that predicted by Th.~9.

Next, let us consider the case that the legitimate users cover the
announced bits by a pre-shared secret key. The covered announced
bits never reduces Eve's Renyi entropy for error correction by
non-interactive linear code. This is because the pre-shared secret
key and the announced bits are mutually independent. Then, the
reduction of the size of the final key by the error correction
process is just as the same as the number of announced bits. In
the case of interactive error correction, however, it is not
apparent whether the announced bits reduce Eve's Renyi entropy or
not, because interactive error correction reveals possible
information on the positions of all errors, which could help Eve.
Thus, it should be verified that this problem is addressed in,
e.g., \cite{lo03}, where the security of CASCADE using the above
method of covering the parity information has been studied.



\end{document}